\documentclass{article}

\usepackage{eusipco}                	

\usepackage[french]{babel}
\usepackage[T1]{fontenc}
\usepackage[latin1]{inputenc}
\usepackage{epsfig}
\usepackage{balance}
\usepackage{amsmath,amstext,amssymb,epsfig,bbm,pslatex,times}

\def\t{\theta}
\def\tc{\hat \theta}

\def\phi{\varphi}

\def\N{\mathbb{N}}
\def\P{\mathbbm{P}}

\def\1{\mathbbm{1}}

\def\ML{\text{ML}}

\def\BIC{\text{BIC}}

\def\CRIT{\text{CRIT}}

\def\ent#1#2{[\![#1,#2]\!]}
\def\qed{\hbox{$\vcenter{\vbox{
   \hrule height 0.4pt\hbox{\vrule width 0.4pt height 6pt
    \kern5pt\vrule width 0.4pt}\hrule height 0.4pt}}$}}

\title{Information criteria and arithmetic codings : an illustration on raw images}

\name {\vspace{0.5cm}}
\name {Guilhem Coq${}^{1}$ , Christian Olivier${}^{2}$ , Olivier Alata${}^{2}$ , Marc Arnaudon${}^{1}$ }

\address {\vspace{1cm}}
\address{  \begin{minipage}[c]{0.50\linewidth} \begin{center}
${}^{1}$ Laboratoire de Mathématiques et Applications \\ 
Université de Poitiers\\ 
Téléport 2 - BP 30179
86962 Chasseneuil FRANCE \\
Phone: +(33) 5 49 49 68 97
Fax: +(33) 5 49 49 69 01 \\ 
email: coq,arnaudon@math.univ-poitiers.fr
          \end{center}\end{minipage}
         \begin{minipage}[c]{0.50\linewidth} \begin{center}
${}^{2}$ Laboratoire Signal Image et Communications\\
Université de Poitiers\\ 
Téléport 2 - BP 30179
86962 Chasseneuil FRANCE \\
Phone: +(33) 5 49 49 65 67
Fax: +(33) 5 49 49 65 70 \\ 
email: olivier,alata@sic.sp2mi.univ-poitiers.fr
          \end{center}\end{minipage}
         }

\begin{document}

\maketitle
\begin{abstract}\begin{em}

In this paper we give a short theoretical description of the general predictive adaptive arithmetic coding technique. The links between this technique and the works of J. Rissanen in the 80's, in particular the BIC information criterion used in parametrical model selection problems, are established. We also design lossless and lossy coding techniques of images. The lossless technique uses a mix between fixed-length coding and arithmetic coding and provides better compression results than those separate methods. That technique is also seen to have an interesting application in the domain of statistics since it gives a data-driven procedure for the non-parametrical histogram selection problem. The lossy technique uses only predictive adaptive arithmetic codes and shows how a good choice of the order of prediction might lead to better results in terms of compression. We illustrate those coding techniques on a raw grayscale image.
\end{em}\end{abstract}

\section{Introduction}

Arithmetic Coding (AC) is an efficient binary coding technique. We use it here in one of its most general form : the \emph{predictive} and \emph{adaptive} one. Even though those aspects of AC are known, it is quite hard to find literature dealing with both of them ; as well as to determine which aspects are actually used in image coding norms such as JPEG and JPEG2000. We try here to answer the first issue but could not collect useful informations about the second. This paper does not seek compression efficiency but wants to show how different AC processes may be used in both parametrical (§\ref{IC}) and non-parametrical (§\ref{lossless}) model selection problems. This explains why we choose to work on raw images. 

After a description of AC algorithm in §\ref{AC}, we take a closer look at the resulting codelength. To this end, we use works of J. Rissanen in \cite{Rissanen_76,Rissanen_86b} and especially \cite{Rissanen_86}. The main conclusion of §\ref{IC} is that the codelength enters the family of information criteria, a widely used tool in the vast problem of model selection. We aim at showing that the \emph{adaptive} aspect of the AC used here is an essential feature.

Next, we design in §\ref{lossless} a new lossless coding technique. It uses a mix between AC, which is compression efficient, and fixed-length coding, which is not. It is shown in §\ref{Choix} that correctly mixing those two methods gives better compression efficiency than using only AC. The most important parameter to be adjusted in order to get that "correct" mix is the order of prediction. Moreover, that method is shown in §\ref{histo} to have a direct application in the histogram selection problem.

Finally we design in §\ref{lossy} a lossy coding technique which, once again, shows the importance of the order of prediction.


\section{Generalities on arithmetic coding}\label{AC}

 \subsection{Multiple Markov Chain} \label{CMM}
 
 The notion of Multiple Markov Chain (MMC) leads to arithmetic coding. Let $E=\{a_1,\dots,a_{m}\}$ be a finite set with $m$ elements. An $E$-valued process $(X_n)_{n \in \N^*}$ is an order $k$ MMC if $k \in \N$ is the smallest integer satisfying the law equality $\P(X_n|X_{n-1},\dots,X_1) = \P(X_n|X_{n-1},\dots,X_{n-k})$ for all $n$. We will always work in the case where that law does not depend on $n$ ; the chain is said homogeneous. An order 0 MMC is a sequence of independent random variables.
 
 If $X$ is an order $k$ MMC, we will suppose that $X_1,\dots,X_k$ are independent and uniformly distributed on $E$. For $i \in E$ a state and $j \in E^k$ a multiple state, we denote by $\t(i|j)$ the probability to see $i$ after $j$. Consequently, choosing $(m-1)m^k$ real numbers $\t(i|j)$ for $j \in E^k$ and $i \in \{a_1,\dots,a_{m-1}\}$ is enough for describing the evolution of $X$. Let $\t$ denote such a parameter and $x^n=x_1,\dots,x_n$ be a sequence of elements of $E$, the likelihood of $x^n$ relatively to $\t$ writes as :
  
 \begin{equation}
 \label{vraisemblance}
 \P(x^n|\t) = \frac{1}{m^k} \prod_{j\in E^k} \prod_{i \in E} \t(i|j)^{n(i|j)} 
 \end{equation}
 where $n(i|j)$ is the number of occurences of $i$ after $j$ in $x^n$.

\subsection{Predictive adaptive arithmetic coding : PAAC} \label{adaptatif}

We deal here with a general AC which is both $k$-predictive and adaptive ; we shorten it to $k$-PAAC. \emph{Predictive} means we code using orders $k$ that may be greater than 1, hence a prediction of the future state of the chain from the current state. \emph{Adaptive} means we do not need any prior knowledge on the chain, except its order ; we learn how to predict the future step by step. Both notions have been formally introduced and studied by Rissanen \cite{Rissanen_76,Rissanen_86b,Rissanen_86}. For a more concrete description of arithmetic coding, we refer to \cite{W_R_N_C} ; note that this paper does not mention the predictive aspect. Let us now give a theoretical description of the general $k$-PAAC algorithm. 

Let $x^n=x_1,\dots,x_n$ be a chain of elements of $E$ to be encoded and $I_c$ be the current interval firstly set to $I_c=[0,1)$. For $n \geq t \geq 1$ we note $x^t = x_1,\dots,x_t$. The only prior we need is an order of coding $k \geq 0$, then the algorithm works as follows.

Suppose that the $t \geq 0$ first symbols are dealt with ; $t=0$ means we have not started the coding yet. To deal with the $(t+1)$-th symbol we actualize transition probabilities as follows :
$$
\tc^{(t)}(i|j) = \frac{ n^{(t)}(i|j) + 1}{n^{(t)}(j) + m}
$$
where $i \in E$, $j \in E^k$, $n^{(t)}(i|j)$ and $n^{(t)}(j)$ denote the respective number of occurences of $i$ after $j$ and of $j$ in the chain~$x^t$ ; $n^{(t)}(j)$ must not count an occurence of $j$ at the very end of that chain. If $k=0$, the multiple states $j$ vanish and we set $n^{(t)}(j)=t$. Those probabilities reflect what we know of the chain at the time $t$ of the coding process; they are the \emph{adaptive} aspect. We then set $j=x_{t-k+1},\dots,x_{t}$ the current state and split the current interval $I_c$ in $m$ smaller intervals according to the probabilities $\tc^{(t)}(i|j)$, $i \in E$. This way, we associate to each possible future state $i \in E$ an interval whose length is proportional to the probability with which we expect it. The $(t+1)$-th symbol is dealt with by choosing for new $I_c$ the interval corresponding to $i=x_{t+1}$.

Once the last symbol $x_n$ has been dealt with, we are left with an interval $I_c=[\hbox{low,high})$. Let $\lceil . \rceil$ denote the superior integer part, there exists two consecutive dyadic numbers with length $\lceil - \log (\hbox{high-low}) \rceil$ in $I_c$. We take as the arithmetic code of $x^n$ the sequence of bits given by the fractionnal part of the biggest one. If encoder and decoder agree on the order $k$ of coding, that sequence of bits is decodable, we refer again to \cite{W_R_N_C}.
 
For illustration in table \ref{ordre_1}, we take $m=2, \ E=\{a,b\}$ and encode $x^4 = abaa$ at order $k=1$. In the splits, we allow the left interval to $a$.

\begin{table}[h]
\caption{Order 1 PAAC of the chain $abaa$.}
\begin{center}
\begin{tabular}{|c|c|c|c|c|}
\hline
$t$  &  $x^t$        &    $I_c$        &   $\hat \t^{(t)}(.|.)$         &     Split \\
\hline
0    &   $\emptyset$ &  $[0,1)$        &  $\begin{array}{c} (a|a)=1/2 \\ (a|b)=1/2 \end{array}$  &   $[0,\frac12,1)$   \\
\hline
1    &   $a$         &  $[0,\frac12)$      &  $\begin{array}{c} (a|a)=1/2 \\ (a|b)=1/2 \end{array}$  &   $[0,\frac14,\frac12)$ \\
\hline
2    &   $ab$        &  $[\frac14,\frac12)$        &  $\begin{array}{c} (a|a)=1/3 \\ (a|b)=1/2 \end{array}$  &   $[\frac14,\frac38,\frac12)$ \\
\hline
3    &   $aba$       &  $[\frac14,\frac38)$     &  $\begin{array}{c} (a|a)=1/3 \\ (a|b)=2/3 \end{array}$  &   $[\frac14,\frac{7}{24},\frac38)$ \\
\hline
4    &   $abaa$      &  $[\frac14,\frac{7}{24})$        &  $\begin{array}{c} (a|a)=\hbox{not used} \\ (a|b)=\hbox{not used} \end{array}$  &   not used \\
\hline
 \multicolumn{5}{|c|}{$\overset{}{\lceil - \log (1/4-7/24) \rceil=5}$} \\
 \multicolumn{5}{|c|}{Code : 01001 ; predecessor : 01000} \\
 \multicolumn{5}{|c|}{Both 1/4+1/32 and 1/4 belong to $I_c$}\\
\hline
\end{tabular}
\end{center}
\label{ordre_1}
\end{table}

This example shows the following general fact about $k$-PAAC : the more unexpected behaviours occur in the chain, the smaller is the last $I_c$, the longer is the code. For instance at step $t=4$ we expected $b$ with probability 2/3, and observed $a$. This caused us to choose the small interval $I_c = [1/4,7/24)$. For comparison, if $b$ had occured the code would have been 0110 which is 1 bit shorter. This leads us to the notion of information criteria (IC).


\section{Information Criteria} \label{IC}

Let us show how the PAAC may be used to solve a model selection problem being : if $x^n$ is a realisation of an unknown MMC (§\ref{CMM}), which is its order ? More precisely, we will see how the \emph{adaptive} aspect of the PAAC is involved.
 
\subsection{Coding approach of the model selection problem}

As mentionned earlier the $k$-PAAC length of $x^n$, say $L(x^n|k)$, is ruled by the unexpected events in $x^n$ : the more unexpected events, the longer the code. Consequently, if $x^n$ is ruled by an unknown order $k^\star$ MMC and we try to $k$-PAAC it at an order $k \neq k^\star$, many unexpected events might occur : either because $k < k^\star$ and we do not look far enough in the past, or because $k > k^\star$ and we take into account informations relative to a too far away past which has actually no influence on the future. Thus the minimization of $L(x^n|k)$ is an appropriate tool for seeking $k^\star$. 

The works of Rissanen will confirm that idea and establish a link with Information Criteria (IC).

\subsection{Rissanen's result}

In \cite{Rissanen_86} it is shown that $L(x^n|k)$ asymptotically behaves as :
\begin{equation}
\label{BIC}
\BIC(x^n|k) = - \log \P(x^n|\tc_k) + \frac{(m-1)m^k}{2} \log n
\end{equation}
where $\tc_k$ is the maximum likelihood (ML) estimator of order $k$ for $x^n$, \emph{i.e.} the parameter that maximizes (\ref{vraisemblance}).

BIC stands for Bayesian Information Criterion and enters the formalism of IC first introduced by Akaike \cite{Akaike_74} ; let us mention \cite{Schwarz,Nishii_88,EMH} in addition to \cite{Akaike_74,Rissanen_86} as important steps in the theory of IC.
 
 Here is the idea behind IC : the first term of the criterion (\ref{BIC}), referred to as the ML term, decreases as $k$ grows. This is mainly because the ML estimator $\tc_k$ fits the datas more accurately if we let him look far away in the past. This phenomena is known as \emph{overparametrization} and is the major problem to be solved in model selection, it appears on figure \ref{fig:superposition}. On the other hand, the second term, the penalty, increases as $k$ grows due to $(m-1)m^k$ which is the number of free parameters in the MMCs model of order $k$. Therefore, the minimization of IC over $k$ realizes a balance between the data fitting, measured by the ML term, and the complexity of the model needed to obtain such a fitting, measured by the penalty. 
 
 The quantity $\BIC(x^n|k)$ is much faster to compute than $L(x^n|k)$ ; the encoder should use BIC before encoding to find which order will achieve the minimum codelength.
 

 One can design a \emph{non-adaptive} order $k$-predictive arithmetic coding process whose codelength would be exactly $\lceil - \log \P(x^n|\tc_k) \rceil = \lceil \ML \rceil$. However, this process requires to send the parameter $\tc_k$ for decodability and, especially, it no longer answers the problem of order selection since ML suffers the overparametrization issue. In terms of IC, the \emph{adaptive} aspect of the process creates the penalty term which avoids overparametrization, see again figure \ref{fig:superposition}.

\subsection{Comparison of actual codings with criterion}

\begin{figure} 
\centerline{\hbox{\psfig{figure=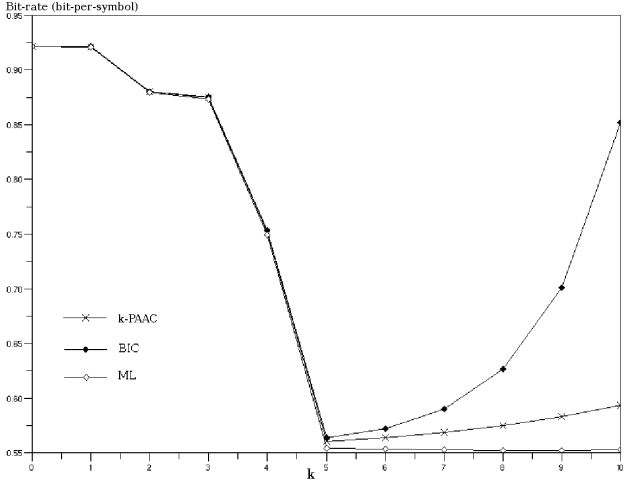,width=8.5cm}}}
\caption{Superposition of codelengths and criteria.}
\label{fig:superposition}
\end{figure}

We generate a realization $x^n$ of an order $k^\star=5$ MMC with $m=2$ and $n=25000$. For $k=0,\dots,10$ we encode it with $k$-PAAC process. We also compute the criterion $\BIC(x^n|k)$ and the quantity $\ML = - \log \P(x^n|\tc_k)$. Results are presented on figure \ref{fig:superposition} divided by $n$ to express them as a bit-rate. 

As expected, BIC and $k$-PAAC curves present a minimum at $k=k^\star$ while the ML method overparametrizes at $k=9$.

Note that, when computing $\BIC$, it is desirable to have enough observations compared to the number of free parameters, empirically : 
\begin{equation}
\label{condition}
n \approx \alpha(m-1)m^k \hbox{ with } \alpha \geq 20
\end{equation}
would be good. If $n$ is too small behind the number of transition probabilities to be estimated, those transitions do not occur often in the chain and their estimation is weak, resulting in the penalty to dominate the ML term. An alternative would be to compute the number of transitions actually observed in the chain and plug them in (\ref{BIC}) instead of $(m-1)m^k$.

\section{Lossless coding of raw images}\label{lossless}

Let $\ent pq$ be the set of integers from $p$ to $q$. Let us choose an $r \times c$ greyscale image and set $n=rc$. Firstly, the image has to be turned into a vector $x^n \in I^n$. For order $k \geq 1$ codings, the way this linearization is done does matter since one does not want to lose proximity information on the pixels. We have chosen the "zigzag" linearization used in $8\times8$ blocks of DCT transform in JPEG norm \cite{Azza_03}. Other transformations have been tested and results are quite similar. Let us now describe our lossless coding method.

\subsection{Lossless coding method}

It is a two-part coding technique. In first, choose a partition $P$ of $I=[0,255]$ ; that is a set of $m$ disjoined intervals $(I_j)_{j\in [\![1,m]\!]}$ whose union is $I$. Then, from $x^n$, form a new chain $y^n$ as follows : 
\begin{equation}
\label{y}
\forall i \in [\![1,n]\!], \ y_i = \sum_{j=1}^m j \1_{I_j}(x_i).
\end{equation}
That is, each $y_i$ denotes the number of the interval of $P$ in which $x_i$ falls. The chain $y^n$ has values in $E=[\![1,m]\!]$. For $k$ an order, we denote by $L(y^n|k,P)$ its $k$-PAAC codelength. If $m=1$, we set $L(y^n|k,P)$ to 0.

Secondly, we denote by $A_j$ the number of integers in $I_j$. Once $y_i=j$ is known one needs, in order to recover $x_i \in I_j$, to specify which one of those integers $x_i$ actually is. This is done for each $x_i \in I_j$ by a simple code with fixed length $\lceil \log A_j \rceil$. Therefore, the number of bits required to recover $x^n$ from $y^n$ is $L(x^n|y^n) = \sum_{j=1}^m n_j \lceil \log A_j \rceil$. 

For decodability, one should also send the partition chosen to encode. We do not take this into account here since the codelength required to this end is very small compared to the quantities $L(y^n|k,P)$ and $L(x^n|y^n)$ we work on.

Let us note $L(x^n|k,P):=L(y^n|k,P) + L(x^n|y^n)$ the total lossless codelength of $x^n$ with help of the partition $P$.

\subsection{Choice of partition and order of prediction}\label{Choix}

As $m$ grows $L(y^n|k,P)$ also grows because $y^n$ has values in $\ent 1m$. By opposition $L(x^n|y^n)$ decreases since the intervals $I_j$ get smaller. Consequently, there should exist a partition $P$ which balances those two phenomena by minimizing the codelentgh $L(x^n|k,P)$. This argument takes place in the theory of Minimum Description Length (MDL) introduced by Rissanen and for which we refer to Grunwald and al. \cite{MDL}.

We estimate $L(y^n|k,P)$ by $\BIC(y^n|k)$, see §\ref{IC}. We then define the following criterion as an estimation of the lossless order $k$ coding of $x^n$ with the partition $P$ : 
\begin{equation}
\label{Critere}
\CRIT(x^n|k,P) = \BIC(y^n|k) + L(x^n|y^n).
\end{equation}

We restrict ourselves to regular partitions ; \emph{i.e.} partitions $P(m)$ whose intervals all have length $256/m$. We work with the $512\times512$ greyscale Lena image. 

\begin{figure} 
\centerline{\hbox{\psfig{figure=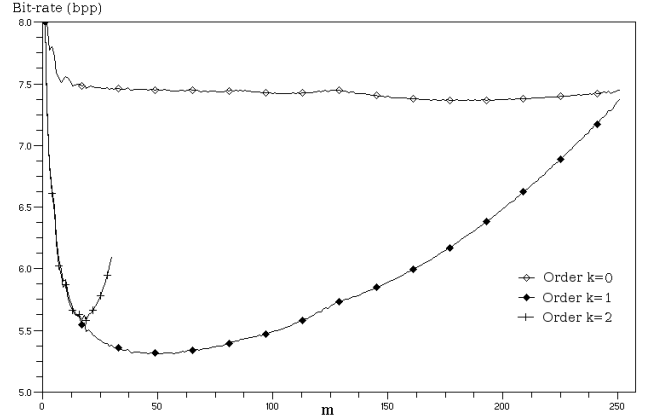,width=8.5cm}}}
\caption{Lossless estimated bit-rates of Lena at order 0,1,2.}
\label{fig:lossless}
\end{figure}

Figure \ref{fig:lossless} presents, for $m$ ranging from 1 to 256 the estimated bit-rate $\CRIT(x^n|k,P(m))/n$ for $k=0,1,2$. For $k=1$, the condition (\ref{condition}) is satisfied for $m$ up to 115 but we still give the $k=1$ curve up to $m=256$ for completeness. The algorithm complexity increases considerably with the order $k$ and computations for $k \geq 2$ shows no significant improvements ; in the case $k=2$ we went up to $m=30$ which makes $\alpha$ about 10. 

 Note that our coding technique with $P(1)$ is equivalent to the pgm format\footnote{http://www.imagemagick.org/script/formats.php}. In the other extreme case, with $P(256)$ we get $y^n=x^n$ and $L(x^n|y^n)=0$ ; this means we directly encode the chain $x^n$ with the $k$-PAAC process. Considering this, figure \ref{fig:lossless} shows how a mix of those two methods leads to better bit-rates. The minimization of the criterion (\ref{Critere}) tells us which partition is to be chosen in order to get the correct mix.
 
 More important, $1$-PAAC is clearly seen to reaches better bit-rates than 0-PAAC : roughly 7 bpp with huge $P(200)$ partition for 0-PAAC against 5.4 bpp with $P(50)$ for 1-PAAC. Note that the order $k$ chosen for the coding process only affects the first term $\BIC(y^n|k)$ of the criterion (\ref{Critere}), hence we may also give the following interpretation of the curves in figure \ref{fig:lossless} : no matter how we quantize them via a partition, the grey scales in our image should not be considered independent but rather of order 1. Unsurprisingsly, that dependance of a pixel greyscale on its neighboors may be shown this way on most of common images which content is comprehensible by the human brain.

\subsection{Histogram selection statistical problem}\label{histo}

It is interesting to note that the criterion (\ref{Critere}) may be directly extended to the histogram selection statistical problem : if $f$ is an unknown density on an interval $I$ and $x^n$ is a sample from this density, which partition of $I$ is to be chosen for building an histogram estimator of $f$ ? 

For such a partition $P$, by independence of $x^n$ and formula (\ref{y}), it is readily seen that the $y_i$'s are independent so that the 0-PAAC of $y^n$ will be the best. Let us denote by $L_j$ the length of $I_j$ and suppose that each $I_j$ contains a number of real numbers proportional to $L_j$. Then, up to terms which do not depend on $P$ and after little calculations, the estimated lossless order 0 codelength of $x^n$ using $P$ is : 
\begin{eqnarray}
\label{Critere_Histo}
\CRIT(x^n|0,P) &=& \BIC(y^n,0) + L(x^n|y^n). \nonumber \\
\CRIT(x^n|0,P) &=& -\sum_{j=1}^m n_j \log \frac {n_j}{nL_j} + \frac{m-1}{2} \log n.
\end{eqnarray}

\begin{figure} 
\centerline{\hbox{\psfig{figure=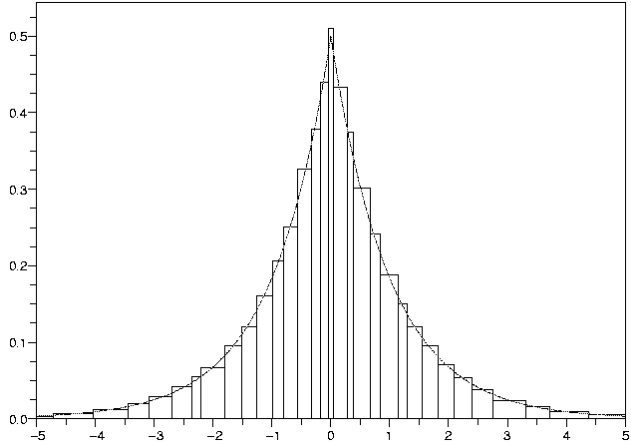,width=8.5cm}}}
\caption{Laplace distribution and histogram chosen by (\ref{Critere_Histo}).}
\label{fig:histo_laplacienne}
\end{figure}

This criterion is in shape really similar to the one used by Birgé and al. in \cite{Birge_06} except it has a coding background which justifies its use. Moreover it is not restricted to regular partitions of $I$. If $I$ is supposed to contain $R$ real numbers, there could be $2^{R-1}$ partitions to be tested, which is huge. Rissanen and al. presented in \cite{Rissanen_92} a dynamic programing method which shrinks to $O(R^2)$ the number of computations required to find which one of the $2^{R-1}$ partitions achieves the minimum of (\ref{Critere_Histo}). For illustration, we present in figure \ref{fig:histo_laplacienne} the partition chosen on a $2000$-sample from the Laplace distribution used to represent DCT coefficients in the JPEG norm. We assume that $I=[-5,5]$ and $R=200$.

\section{Lossy coding of raw images}\label{lossy}
We keep the same linearization as in §\ref{lossless} to turn an image into a vector $x^n$ and now describe our lossy coding method.

\subsection{Lossy coding method}

For $P$ a partition of $[0,255]$ in $m$ intervals, we define the $[\![1,m]\!]$-valued chain $y^n$ as in (\ref{y}). Next, we quantize the datas $x^n$ on $P$ at their barycenter. That is, for each $j \in [\![1,m]\!]$, we consider all $x_i$'s falling into $I_j$, compute their barycenter, round it to the closest integer $B_j$ and finally set all those $x_i$'s to $B_j$. This gives a new image with only $m$ grey levels, this is where the loss occurs. Moreover, that quantization creates an injective map :
\begin{equation*}
\begin{array}{cccc}
B: & [\![1,m]\!] & \longrightarrow & [\![0,255]\!] \\
   &      j      & \longmapsto      &   B_j
\end{array}
\end{equation*}
With the help that map, the decoder is able to reconstruct the quantized image from only the chain $y^n$ ; therefore $B$ is to be sent. However, the coding of such a map is very short compared to the codelength of the chain $y^n$, so we drop it.

Now we are left to encode $y^n$ with the $k$-PAAC process, hence the estimation of the lossy codelength of our image by the BIC criterion (\ref{BIC}) :
$$
\BIC(y^n|k) = - \log \P(y^n|\tc_k) + \frac{(m-1)m^k}{2} \log n.
$$

\subsection{Influence of the order on bit-rates}

\begin{figure} 
\centerline{\hbox{\psfig{figure=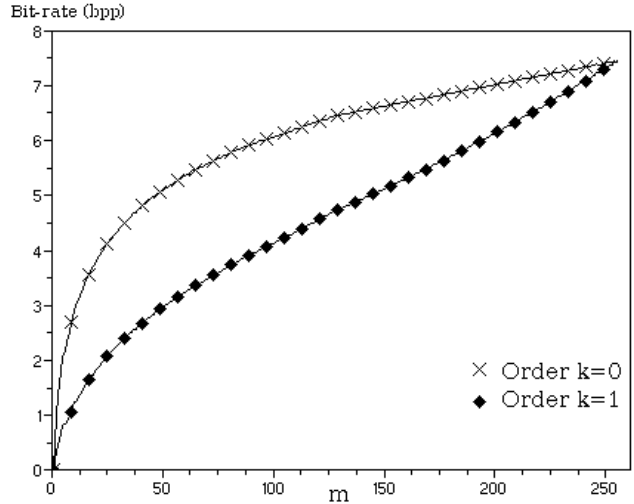,width=8.5cm}}}
\caption{Estimated Lena's bit-rates for 0-PAAC and 1-PAAC.}
\label{fig:debit}
\end{figure}

We still restrict ourselves to regular partition $P(m)$ and work with Lena. Figure \ref{fig:debit} presents the estimated bit-rates $\BIC(y^n|k)/n$ for $m$ ranging from 1 to 256 and orders $k=0,1$. For any $m$, the fact that the $k=1$ curve is under the $k=0$ curve means, as in §\ref{lossless} and via IC interpretation, that the chain $y^n$ is of order 1 rather than order 0.

\subsection{Comparison involving distortion}

\begin{figure}[h] 
\centerline{\hbox{\psfig{figure=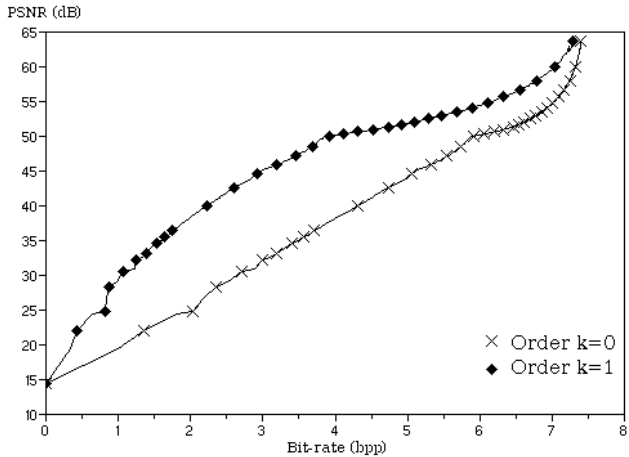,width=8.5cm}}}
\caption{Estimated Lena's bit-rates/PSNR for 0 and 1-PAAC.}
\label{fig:dd}
\end{figure}

\begin{figure}[h] 
\begin{center}
\begin{tabular}{cc}
\psfig{figure=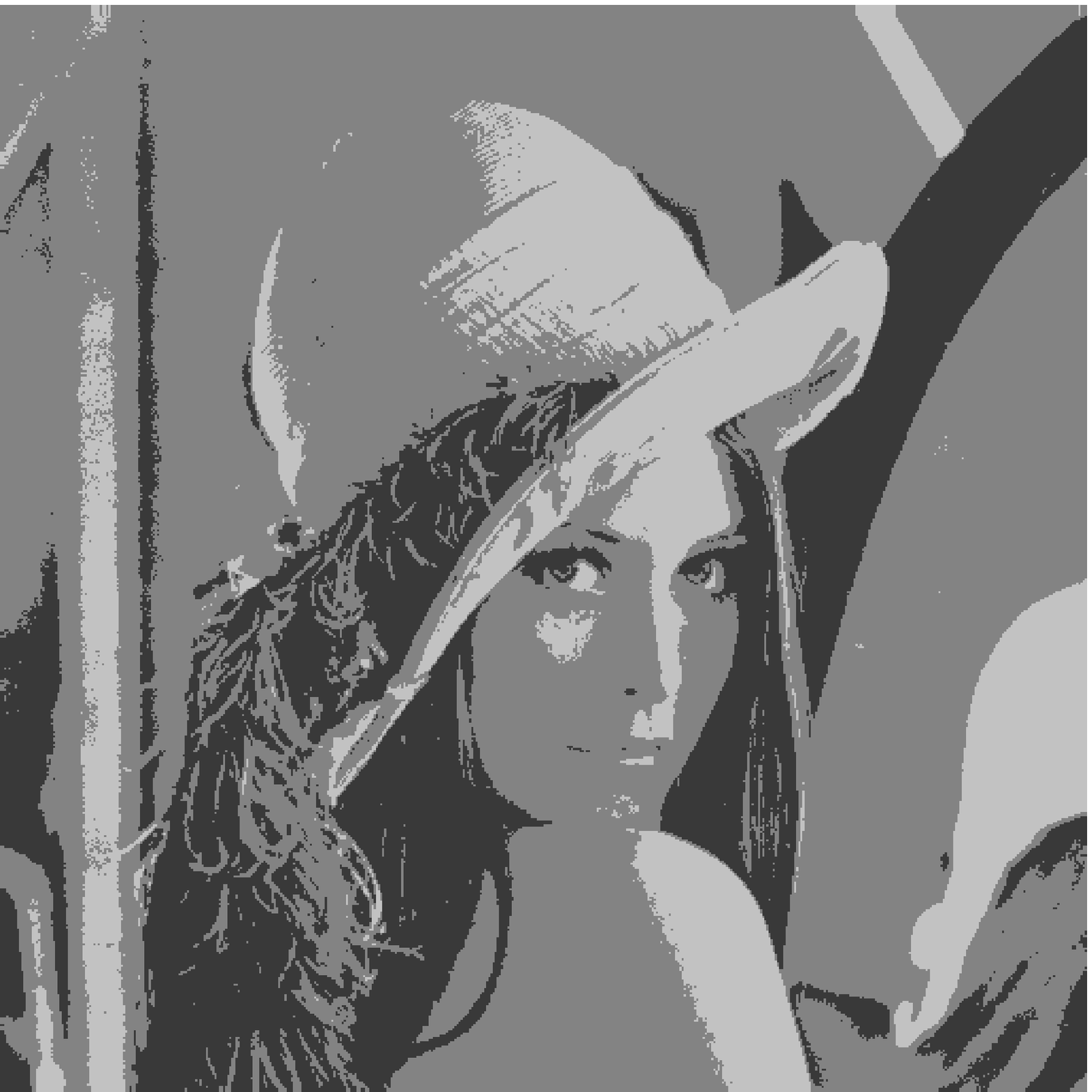,width=4cm} & 
\psfig{figure=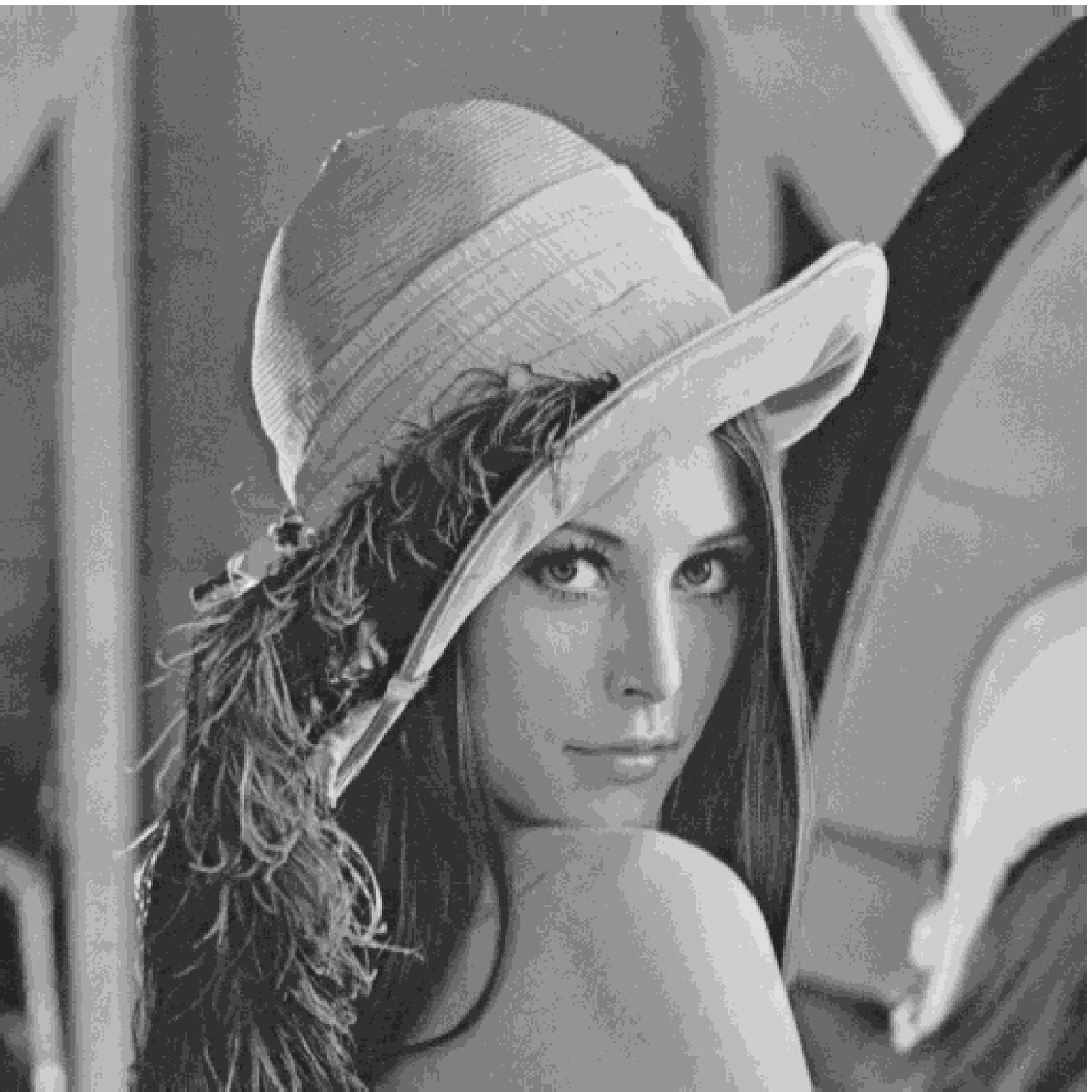,width=4cm} \\
  $m=3$ levels : 22.11 dB & $m=13$ levels : 33.15 dB \\
 0-PAAC : 1.36 bpp   & 0-PAAC : 3.18 bpp    \\
 1-PAAC : 0.43 bpp   & 1-PAAC : 1.39 bpp    \\
\end{tabular}
\end{center}
\caption{Estimated PSNR and bit-rates on Lena quantized at $m=3$ and $m=13$ levels for 0-PAAC and 1-PAAC.}
\label{fig:Lenas}
\end{figure}

Each value of $m$ brings a certain quantization, thus a certain distortion. We measure this distortion by the Peak Signal to Noise Ratio (PSNR) and plot it against the corresponding bit-rate of 0-PAAC and 1-PAAC in figure \ref{fig:dd}. For illustration, we present in figure \ref{fig:Lenas} the two quantized Lena images obtained for $m=3$ and $m=13$ with their respective PSNR. We also give bit-rates achieved by 0-PAAC and 1-PAAC on each of those image. For instance, this shows that at an imposed rate of about 1.4 bpp, the 1-PAAC allows to encode Lena with a PSNR of 33.15 dB while the 0-PAAC only gives 22.11 dB.

\section{Perspectives}

As mentionned in the introduction we did not provide efficient compression results by intentionally working on raw images. Therefore it would be interesting to insert the discussed binary coding methods after, for instance, the wavelet transform block of the JPEG2000 norm. In order to compress, one should in first determine with the BIC criterion (\ref{BIC}) the order of the sequence of wavelet coefficients and then use the criterion (\ref{Critere}) to determine the partition which allows to encode those coefficients efficiently.

\newpage

\begin{center}
{\bf ACKNOWLEDGMENTS}
\end{center}
The authors would like to thank the PIMHAI, INTERREG IIIB "Arc Atlantique" project for its support in the writing of this paper. 

First published in \emph{Proceedings og the 15th European Signal processing Conference EUSIPCO 2007} in 2007, published by EURASIP.

\end{document}